\pdfoutput=1

\documentclass[useAMS,usenatbib,fleqn]{mnras}

\usepackage{graphicx}
\usepackage{amsmath,amssymb}
\usepackage{natbib}

\title[Heteroclinic connections in barred galaxies]{Orbital and escape dynamics in barred galaxies - IV. \\ Heteroclinic connections}

\author[E. E. Zotos \& Ch. Jung]{Euaggelos E. Zotos$^1$\thanks{E-mail: evzotos@physics.auth.gr} and Christof Jung$^2$\thanks{E-mail: jung@fis.unam.mx} \\
$^1$ Department of Physics, School of Science, Aristotle University of Thessaloniki, GR-541 24, Thessaloniki, Greece \\
$^2$ Instituto de Ciencias F\'{i}sicas, Universidad Nacional Aut\'{o}noma de M\'{e}xico Av. Universidad s/n, 62251 Cuernavaca, Mexico
}

\begin{document}

\date{Accepted 2019 May 9. Received 2019 May 3; in original form 2019 January 23}

\pubyear{2019} \volume{487} \pagerange{1233--1247}

\setcounter{page}{1233}

\maketitle

\label{firstpage}

\begin{abstract}
Continuing the series of papers on a new model for a barred galaxy, we investigate the heteroclinic connections between the two normally hyperbolic invariant manifolds sitting over the two index-1 saddle points of the effective potential. The heteroclinic trajectories and the nearby periodic orbits of similar shape populate the bar region of the galaxy and a neighbourhood of its nucleus. Thereby we see a direct relation between the important structures of the interior region of the galaxy and the projection of the heteroclinic tangle into the position space. As a side result, we obtain a detailed picture of the primary heteroclinic intersection surface in the phase space.
\end{abstract}

\begin{keywords}
stellar dynamics -- galaxies: kinematics and dynamics -- galaxies: spiral -- galaxies: structure
\end{keywords}

\defcitealias{JZ16a}{Part I}
\defcitealias{JZ16b}{Part II}
\defcitealias{ZJ18}{Part III}

\section{Introduction}
\label{intro}

In disc galaxies, which contain a rotating bar, the index-1 saddle points $L_2$ and $L_3$ are very important for the whole dynamics of the galactic system. These Lagrange points are directly associated with the corresponding Lyapunov orbits \citep{L07,L49} and the respective normally hyperbolic invariant manifolds (NHIMs). We can imagine the stable and unstable manifolds of the NHIMs as tubes inside the phase space, which guide and control the motion of stars through the Lagrange points $L_2$ and $L_3$. Therefore, the NHIMs are very important for the escape dynamics of barred galaxies. Moreover, the manifolds are also related with the observed stellar structures, such as rings and spirals, in galaxies with a bar. \citet{RGMA06} discussed how the manifolds affect the shape and the velocity of rings. In the same vein, the analysis was expanded in a series of papers in an attempt to determine the correlations between the manifolds and the rings and spirals in barred galaxies \citep{RGAM07,ARGM09,ARGBM09}, while a comparison with related observational data has been performed in \citet{ARGBM10}. In another series of papers, $N$-body simulations revealed the role of the manifolds in the observed stellar structures \citep{VTE06}, the effect of ``stickiness", which slows down the rate of escape \citep{TEV08}, and the role of non-axisymmetric components \citep{TKEC09}.

For many years, the Ferrers' triaxial model \citep{F77} was the only realistic model for describing the motion of stars in barred galaxies. However, the main disadvantage of this model is its high mathematical complexity, regarding the corresponding potential (which is not known in closed form) and the equation of motion \citep{P84}. On this basis, in \citet{JZ15} we introduced a new barred galaxy model with a much simpler bar potential, which requires significantly less computational time compared to the Ferrers' potential.

Our model potential for a single barred galaxy has been introduced and explained in all details in \citet{JZ15} and used in \citet{JZ16a} (hereafter \citetalias{JZ16a}), \citet{JZ16b} (hereafter \citetalias{JZ16b}) and \citet{ZJ18} (hereafter \citetalias{ZJ18}). Therefore, mainly for saving space, we do not repeat the presentation of the model, we just give some short remarks on its important properties: The total gravitational potential consists of four parts which describe the nucleus, the bar, the disc and the halo, respectively. Because we use a description of the dynamics in a rotating frame of reference, the effective potential consists of the sum of the total gravitational potential and the centrifugal potential. We use a coordinate system where the plane of the disk lies in the $(x,y)$ plane. All the saddle points of the effective potential lie in this plane $z=0$. A plot of the effective potential in this horizontal plane has been given in Fig.~1 in \citetalias{JZ16a}. The most important saddle points of index-1 are the Lagrange points $L_2$ and $L_3$. The numerical parameter values in the potential are chosen with the galaxy NGC 1300 in mind.

Over the saddle points of index-1 we expect to find normally hyperbolic invariant manifolds of codimension 2 (see also the detailed explanation in section 4 of \citetalias{JZ16b}). More details regarding the NHIMs can be found in \citet{W94}. These NHIMs have stable manifolds and unstable manifolds of codimension 1 which direct and channel the global behaviour of the dynamics of the whole system to a large extent. From each saddle NHIM there is a branch of its stable manifold and a branch of its unstable manifold going to the outside and also another branch of its stable manifold and another branch of its unstable manifold going to the inside. A main topic of \citetalias{JZ16b} was, to show that the unstable manifolds going to the outside determine the structure of rings or spirals of the galaxy. Now we show the role of the branches of the stable and unstable manifolds going to the inside. The main direction of arguments will be to show how these inner branches and their heteroclinic connections are related to the shape of the nucleus and of the bar of the galaxy.

The Lyapunov orbits, over a saddle point, are the most important periodic orbits within the corresponding saddle NHIM. In our case of a 3 degrees of freedom (3-dof) system we have one horizontal Lyapunov orbit (in the following called $l_h$) and one vertical Lyapunov orbit (in the following called $l_v$) over each one of the index-1 saddles. These particular periodic orbits and their development scenarios as a function of the energy have been described in detail in subsection 4.3 of \citetalias{JZ16b} and plots of these orbits in the position space have been given in Fig.~6 of \citetalias{JZ16b}.

Of course, it should be clear that the NHIMs are of essential importance for the dynamics only for energies close to and slightly above the saddle energy. For such energies the NHIMs direct the complete escape processes and they fix also the global dynamics and the structures formed in the system to a large extent. For the parameter case used mainly in previous publications and also here (named the standard model) the saddle energy is $E_s = -3242$. In the present text we will restrict all considerations to the energy value $E = -3200$. This is still an energy typical for the escape processes and we find a qualitatively equal behaviour for all other energy values, a little above the saddle energy.

\begin{figure}
\begin{center}
\includegraphics[width=\hsize]{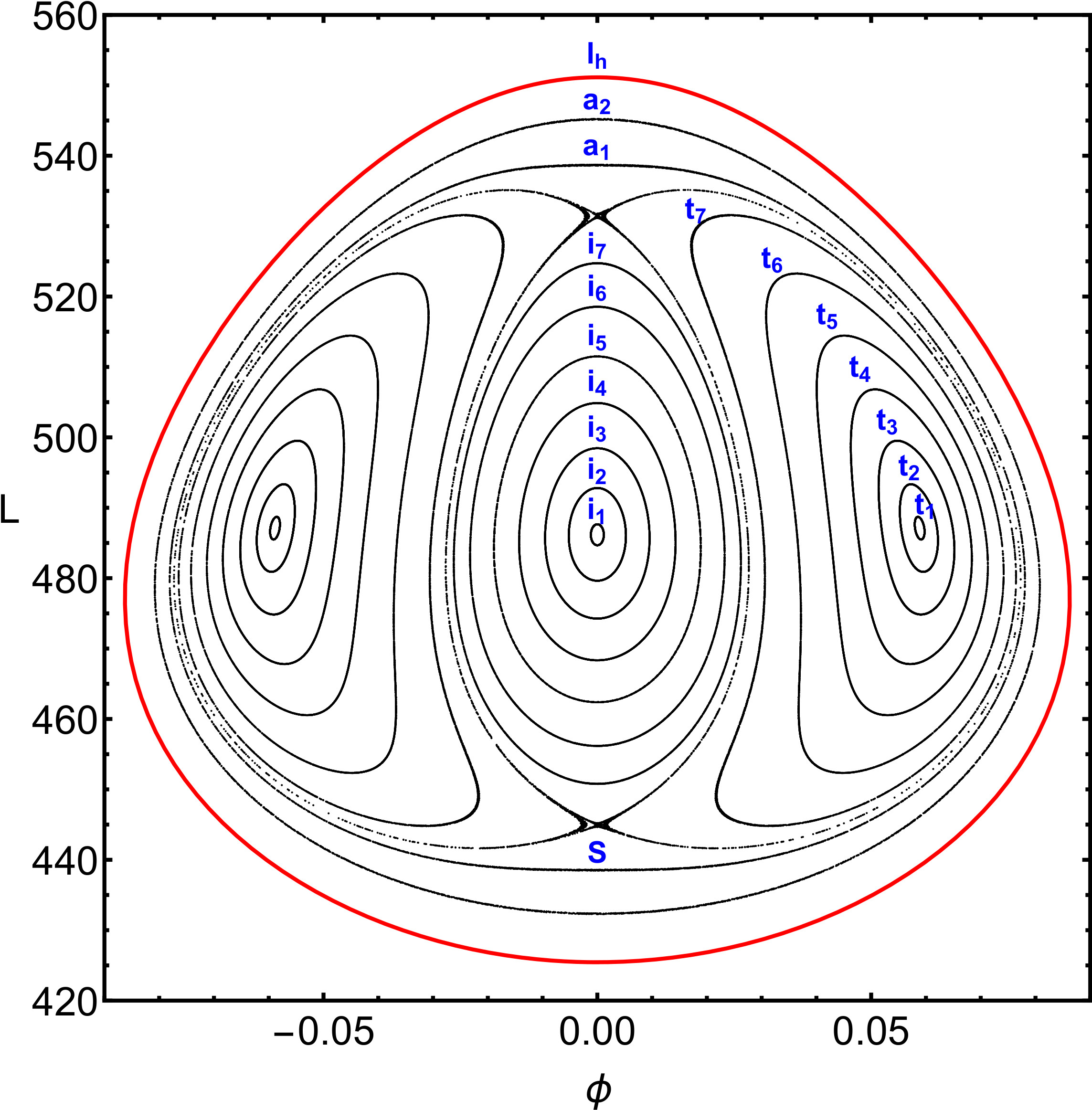}
\end{center}
\caption{Plot of the restricted Poincar\'e map on the NHIM in the coordinates $\phi = \arctan(y/x)$ and $L = x p_y - y p_x$. Many iterations of a moderate number of initial points are plotted. The structures belonging to the various initial points are labelled. The red boundary curve is $l_h$.}
\label{map}
\end{figure}

In the following, we rely on some properties of the saddle NHIMs which are important for the argumentation: The NHIMs are invariant and therefore the restriction of the Poincar\'e map to the NHIMs exists, we call it the restricted map $M_{\rm res}$ (for details on how to construct this $M_{\rm res}$ see \citet{GDJ14}). As in the \citetalias{JZ16b} and \citetalias{ZJ18}, we use $z = 0$ as intersection condition for the Poincar\'e map. Because of symmetry reasons the intersection orientation is irrelevant. For a 3-dof system the $M_{\rm res}$ acts on a 2-dimensional domain and is very similar to a usual Poincar\'e map for a 2-dof system, it is the Poincar\'e map for the internal 2-dof dynamics of the NHIM. Therefore it is the ideal graphical representation for this internal dynamics of the NHIMs. In \citetalias{JZ16b} the whole development scenario of the $M_{\rm res}$, as function of the energy, has been presented and discussed in detail, see Figs.~7 and 8 in \citetalias{JZ16b}. Because in the present text we restrict all considerations to the energy level -3200 we repeat in Fig.~\ref{map} the restricted map for just this single energy value.

The plot is represented in the canonical coordinates $\phi = \arctan(y/x)$ and $L = x p_y - y p_x$. As usual, for the Poincar\'e maps we show many iterations under $M_{res}$ of a moderate number of initial points. For more detailed explanations see subsection 4.4 of \citetalias{JZ16b}. In the following, the words tangential and normal always refer to directions relative to the NHIM surface.

In Fig.~\ref{map} note the following properties: We are still close to the saddle energy, therefore the map looks rather regular, it is still close to an integrable map, there are no large scale chaos regions. The fixed point in the middle represents $l_v$, while the boundary represents $l_h$. The fixed points at $\phi \approx \pm 0.06$ and $L \approx 486$ represent a pair of tilted loop orbits split off from $l_v$ at $E \approx -3223$. These orbits are tangentially stable. At the energy $E \approx -3214$ $l_v$ splits off another pair of tilted loop orbits, they are tangentially unstable and are represented in the $M_{\rm res}$ as the centres of a fine chaos strip which appears in the diagram like a separatrix. This separatrix separates 3 systems of concentric KAM curves: First, the curves around the central fixed point. Second, the curves around the tilted loop orbits. Note that because of the $z$ reflection symmetry the two tilted loop islands can be identified and treated as a single island structure. And as a third set of curves we have the curves running parallel and near to the boundary.

Later we will refer to the individual curves seen in Fig.~\ref{map} and we give these curves the following names also included in the figure: The curves in the inner island are called $i_1$ to $i_7$, from the inside counting outwards. The curves in the tilted loop islands are called $t_1$ to $t_7$ again outgoing from the centre to the outside and the curves running parallel to the boundary are called a1 and a2, again counting outwards.

\section{Periodic orbits approaching heteroclinic connections}
\label{per}

\begin{figure*}
\centering
\resizebox{\hsize}{!}{\includegraphics{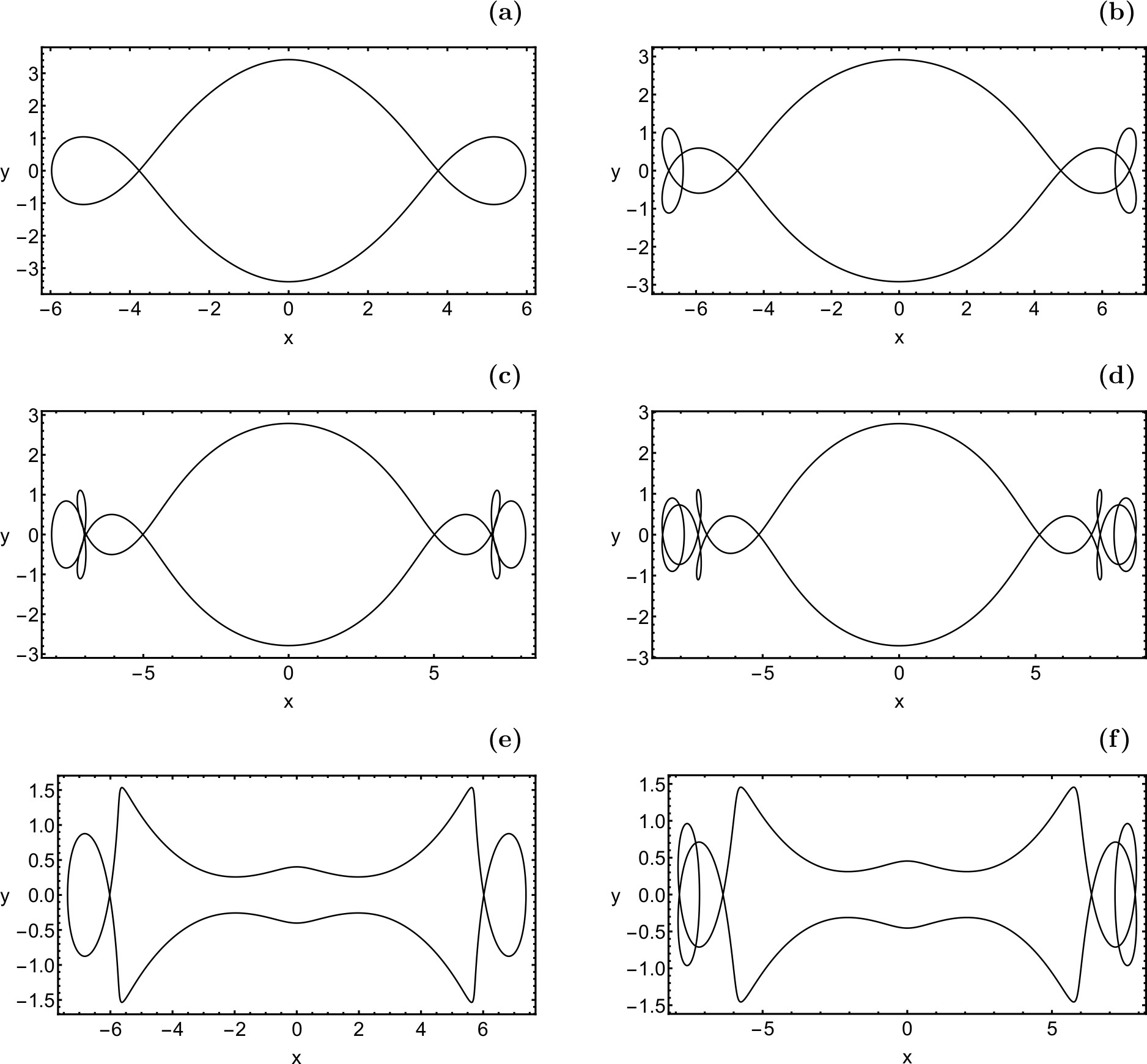}}
\caption{The various panels show 6 horizontal periodic orbits on the $(x,y)$ plane. The $x:y$ resonance ratios are 1:3, 1:5, 1:7, 1:9, 1:3, 1:5, respectively.}
\label{orbs}
\end{figure*}

The two NHIMs of codimension 2 sit over the index-1 saddle points $L_2$ and $L_3$ of the effective potential. In the following we call NHIM2 the NHIM over $L_2$ and NHIM3 the NHIM over $L_3$. The inner branches of their stable and unstable manifolds run into the region of the bar and they form heteroclinic intersections. The corresponding heteroclinic trajectories start on one NHIM and end on the other one. We can even do more. We can look for heteroclinic connections not just between the NHIMs, we can look for heteroclinic connections between individual substructures identified in the two NHIMs. For stable and unstable manifolds of NHIMs a foliation theorem holds (see chapter 5 in \citet{W94}) which shows that the internal structures of the NHIMs are transported along these manifolds.

\begin{figure*}
\centering
\resizebox{\hsize}{!}{\includegraphics{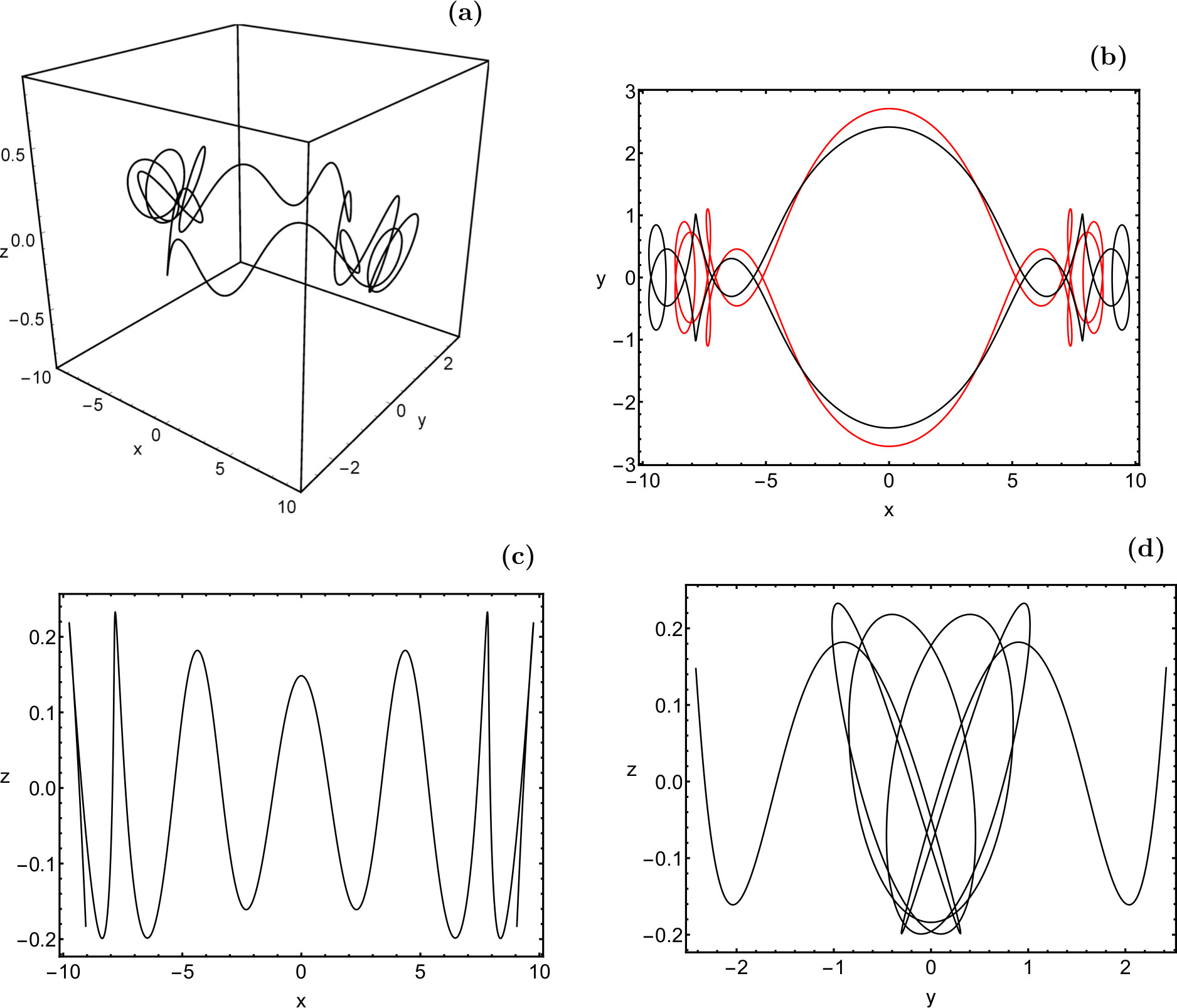}}
\caption{A symmetric periodic orbit (class 2) with resonance ratio $x:y:z = 1:9:14$. Part (a) is a perspective view in the 3 dimensional position space $(x,y,z)$. Parts (b), (c), and (d) are the projections into the various 2 dimensional coordinate planes. For comparison, in part (b) also the horizontal orbit from Fig.~\ref{orbs}d has been included in red. (Colour figure online).}
\label{orb1}
\end{figure*}

\begin{figure*}
\centering
\resizebox{\hsize}{!}{\includegraphics{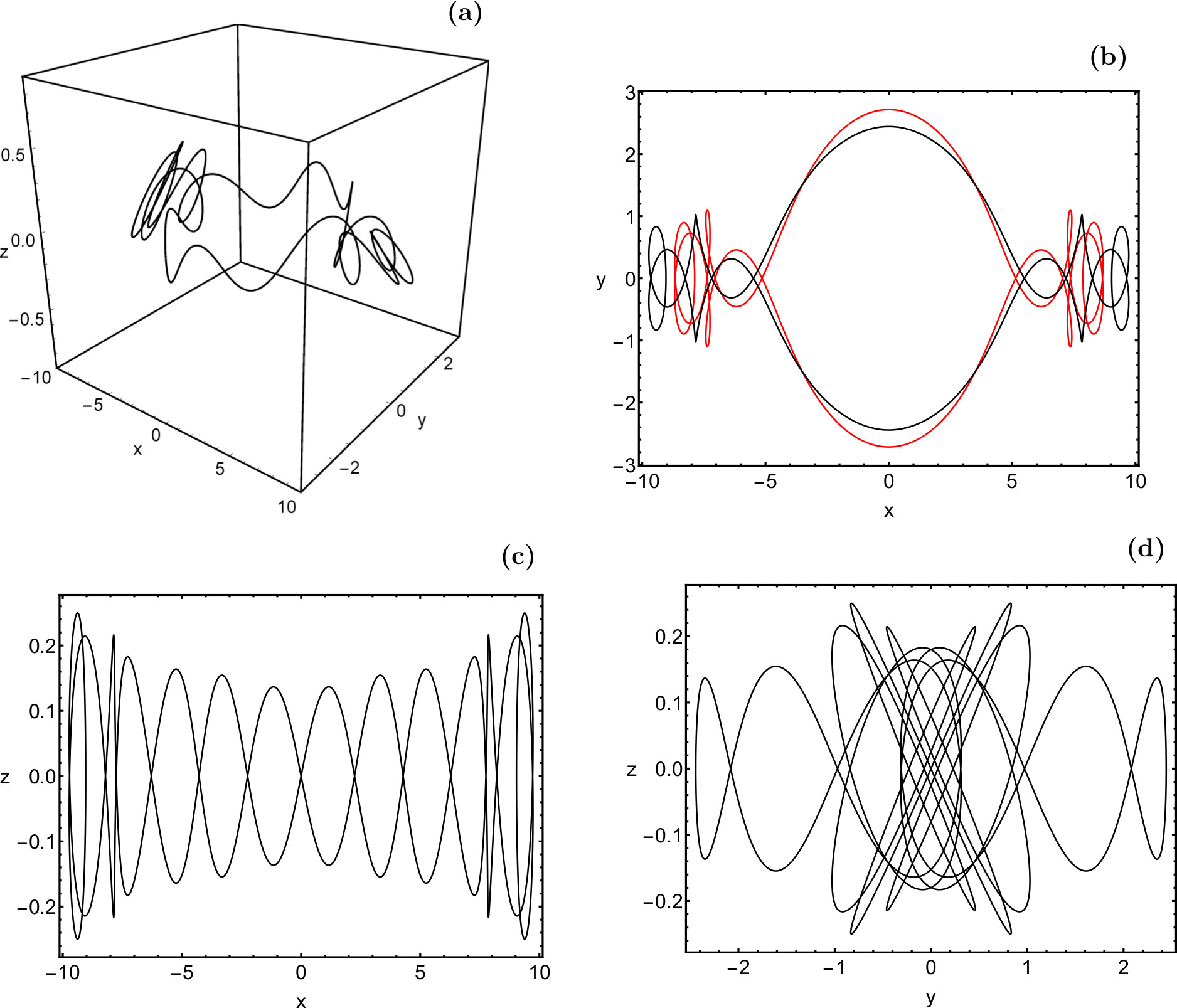}}
\caption{An antisymmetric periodic orbit (class 3) with resonance ratio $x:y:z = 1:9:14$. Part (a) is a perspective view in the 3 dimensional position space $(x,y,z)$. Parts (b), (c), and (d) are the projections into the various 2 dimensional coordinate planes. For comparison, in part (b) also the horizontal orbit from Fig.~\ref{orbs}d has been included in red. (Colour figure online).}
\label{orb2}
\end{figure*}

In addition to the heteroclinic connections themselves we study the periodic orbits running in the neighbourhood of heteroclinic connections. Remember that heteroclinic trajectories are accumulation points of periodic orbits. They are periodic orbits which oscillate between the neighbourhoods of NHIM2 and NHIM3, i.e. between the two saddle points $L_2$ and $L_3$ of the effective potential. Because of discrete symmetry NHIM3 is obtained from NHIM2 by a rotation of the system around the $z$-axis by an angle $\pi$. Let us call this symmetry operation $D_z(\pi)$ in the following. This discrete symmetry leads to a corresponding symmetry in many heteroclinic structures and in many periodic orbits and we will exploit this symmetry whenever we can do so.

The system has another important discrete symmetry which will also be useful. It is the reflection symmetry in $z$ direction. These two discrete symmetries together lead to the existence of the following classes of most basic, most simple periodic orbits and also to corresponding classes of heteroclinic trajectories. We call a periodic orbit simple if it intersects the plane $x = 0$ only once in each one of the two orientations, during one period. This implies that in a classification of these orbits by resonance relations in the 3 degrees of freedom (the 3 coordinate directions) the resonance number of $x$ for simple periodic orbits is always 1. Of course, in addition there are also periodic orbits and heteroclinic trajectories without discrete symmetry and with various intersection numbers of the plane $x = 0$ in each one of the two orientations. But the simple symmetric ones are also the shortest ones and it makes sense to study first and mainly the simple ones having some additional symmetry properties, with respect to the $z$ reflection. First, there is a class of periodic orbits which we will call the horizontal class 1: These orbits always have $z \equiv 0$ and $ p_z \equiv 0$ and in addition as a point set (not taking care of the orientation of motion) these orbits are symmetric under $x$ reflection and under $y$ reflection.

Fig.~\ref{orbs} shows some periodic orbits of this class 1. We observe that parts (a), (b), (c) and (d) represent the $x : y$ resonances 1:3, 1:5, 1:7, 1:9, respectively of a sequence of orbits which are relatively wide at their moment of crossing the line $x = 0$. Of course, also the corresponding continuations of this sequence, with resonances $x:y=1:n$ for all larger odd integers $n$ exist. It should be obvious from the plot how this sequence converges to a horizontal heteroclinic trajectory going from $L_2$ to $L_3$ and its $D_z(\pi)$ rotated counterpart which goes from $L_3$ to $L_2$. Parts (e) and (f) are the beginning of another sequence which is rather narrow at the moment of the crossing of the line $x = 0$. The 1:3 and the 1:5 resonances are plotted. Also this sequence has its continuation and converges against another horizontal heteroclinic trajectory. More on the limiting horizontal heteroclinic trajectories comes below.

Next let us consider periodic orbits which also perform motion in the $z$ direction. The simplest ones are periodic orbits whose projection into the horizontal plane are qualitatively equal to the horizontal orbits of Fig.~\ref{orbs} and of the continuation of this sequence. With respect to the phase relation between the horizontal motion and the $z$ motion we find two particularly simple possibilities. First, we can have that $p_z = 0$ when the trajectories cross the plane $x = 0$. We call such orbits symmetric $z$ excitations and put them into class 2. Under a $x$ reflection the point set of these orbits is $z$ invariant. Second, we can have that $z = 0$ when the orbit crosses the plane $x = 0$. We call such orbits antisymmetric $z$ excitations and put them into class 3. Under $x$ reflection the point set of these orbits is also $z$ reflected.

In Fig.~\ref{orb1} we present an orbit of class 2 (perspective view in part (a) and the projections into the three coordinate planes $(x,y)$, $(x,z)$ and $(y,z)$ in parts (b), (c) and (d), respectively), it shows a $x:y = 1:9$ resonance in its horizontal projection and this horizontal projection is qualitatively similar to the periodic orbit in Fig.~\ref{orbs}d, i.e. to the one representing the $x:y = 1:9$ resonance. For comparison, this horizontal orbit is also included in part (b)
in red colour. In the 3 dimensional position space the orbit from Fig.~\ref{orb1} performs the $x:y:z$ resonance $1:9:14$.

Fig.~\ref{orb2} presents the corresponding orbit of class 3 with the same resonance ratio $x:y:z = 1:9:14$. Again, for comparison the horizontal 1:9 orbit is included in part (b) in red colour. To each periodic orbit of class 2 and of class 3 exists also the $z$ reflected periodic orbit.

Panel (d) of Fig.~\ref{orb2} helps us to make the following comment evident. We see trajectory segments running close to the $y - z$ diagonal and the corresponding antidiagonal. Clearly, along the diagonal or the antidiagonal the $y$ and $z$ degrees of freedom run in a 1:1 resonance. These trajectory segments are the ones running in the outer parts of the bar. We also see a trajectory segment running in an approximate $y:z = 1:5$ resonance. It is the segment coming from the inner part of the bar, where the trajectory makes a large semi-loop around the nucleus. And the total $y:z$ resonance ratio depends on the relative length of the time
intervals in which the orbit runs in the approximate 1:1 resonance and the one in which it runs in the approximate 1:5 resonance. In the particular example shown in this figure this time ratio happens to turn out such that the resulting total resonance ratio becomes $y:z = 9:14$.

It is obvious how the logical continuation of the sequence of orbits presented in Fig.~\ref{orbs} approaches a heteroclinic connection. With increasing $y$ number in their resonance relation these orbits come closer to the saddles and spend more time near the saddles making more loops in the saddle region. In the limit the time over the saddles diverges to infinity and thereby the limit of these orbits turns into horizontal heteroclinic connections. The corresponding sequences of periodic orbits with $z$ excitation converge to heteroclinic trajectories with $z$ excitation.

\section{Heteroclinic trajectories}
\label{het}

When we are looking for simple heteroclinic trajectories then we can again look first at a class 1, which contains horizontal trajectories, i.e. trajectories which lie completely on the horizontal $(x,y)$ plane. They are the most simple and most symmetric heteroclinic connections between $l_h(L_2)$ and $l_h(L_3)$.

\begin{figure}
\begin{center}
\includegraphics[width=\hsize]{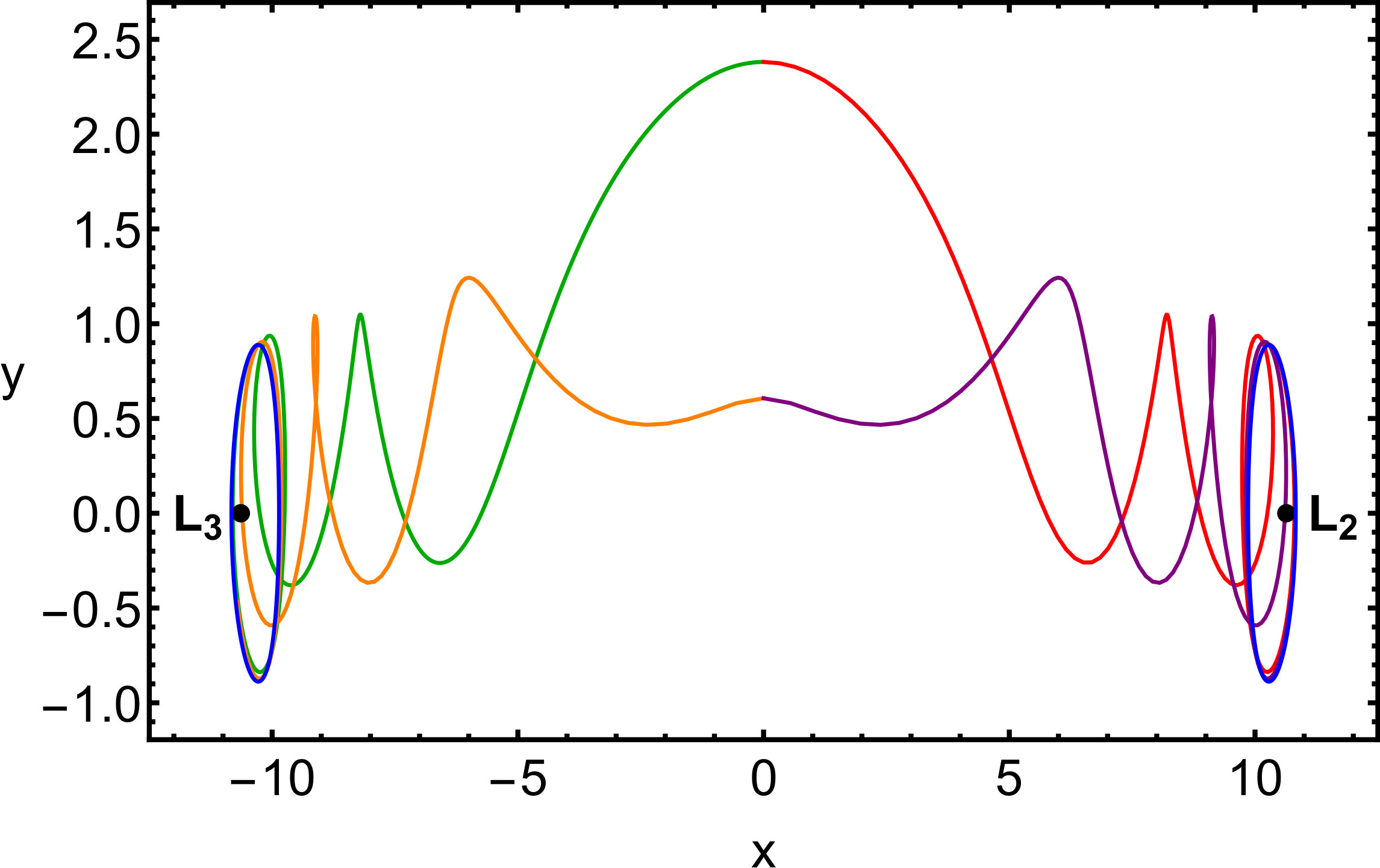}
\end{center}
\caption{The two simple horizontal heteroclinic trajectories connecting $l_h(L_2)$ and $l_h(L_3)$. The parts of the trajectories with $x > 0$ (past parts) are plotted in red and in purple, respectively. The parts with $x < 0$ (future parts) are plotted in green and in orange, respectively. Also included in blue colour are $l_h(L_2)$ and $l_h(L_3)$. The saddle points $L_2$ and $L_3$ themselves are included as black dots. (Colour figure online).}
\label{pn}
\end{figure}

There are two trajectories of this type and they are presented in Fig.~\ref{pn}. They both start in the past near $l_h(L_2)$ and end in the future near $l_h(L_3)$. Of course, there also exist the two corresponding heteroclinic trajectories going from $L_3$ to $L_2$. They are obtained by an application of $D_z(\pi)$ to Fig.~\ref{pn}. Note that each one of the heteroclinic trajectories from Fig.~\ref{pn} taken together with its rotated counterpart has the same symmetry properties as each one of the periodic orbits from Fig.~\ref{orbs}. In analogy to the periodic orbits we call a heteroclitic trajectory simple if it intersects the plane $x = 0$ only once, it does it in negative orientation if it goes from NHIM2 to NHIM3 and it does it in positive orientation if it goes from NHIM3 to NHIM2. As initial
conditions for the two heteroclinic trajectories we take their point of intersection with the line $x = 0$ and integrate forward (green and orange colour in the plot) and backward (red and purple colour in the plot). The limit sets, namely $l_h$ over the saddles, are also included in blue colour.

\begin{figure*}
\centering
\resizebox{\hsize}{!}{\includegraphics{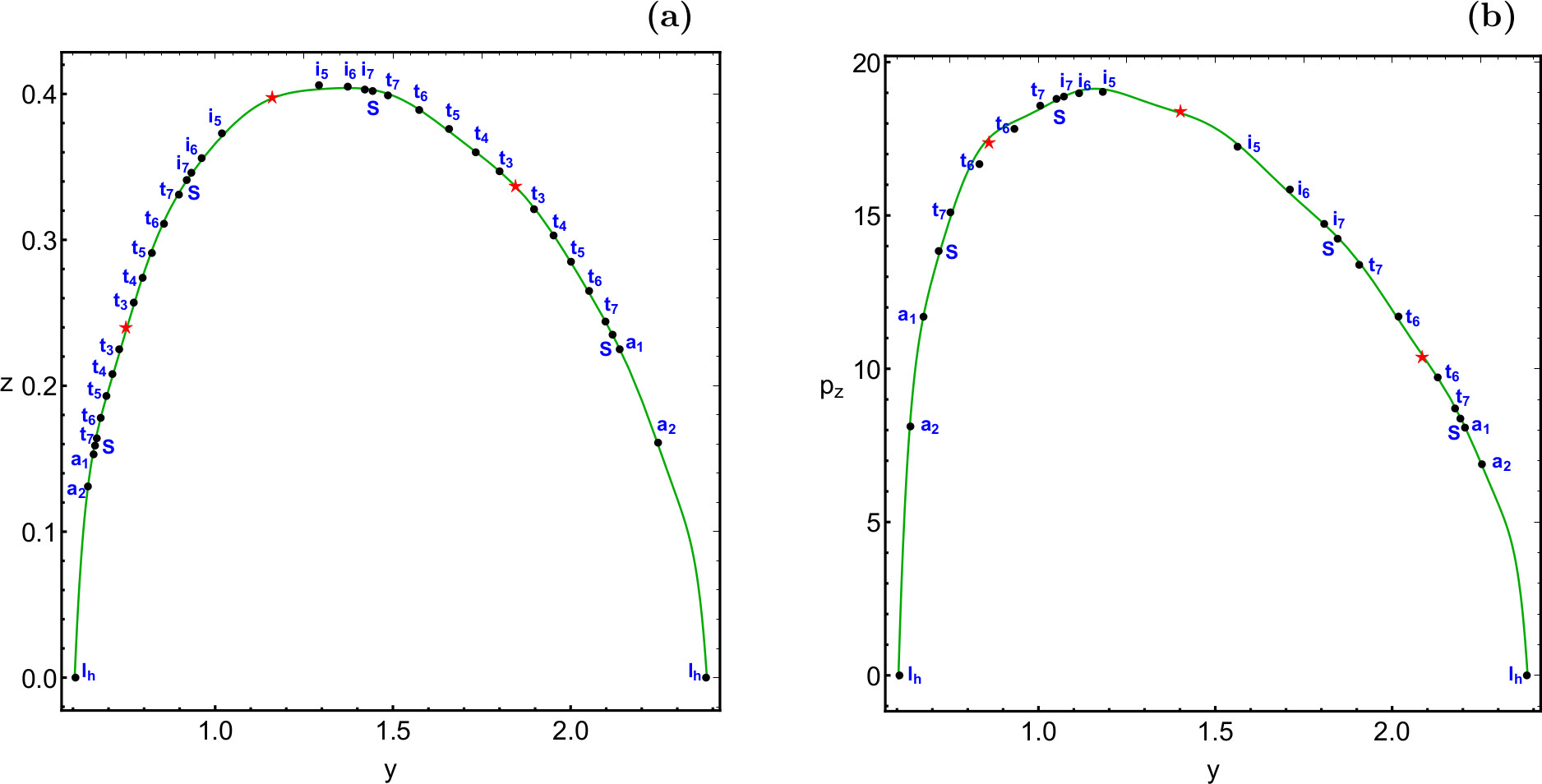}}
\caption{The intersection coordinates with the plane $x = 0$ and negative orientation of the simple symmetric heteroclinic trajectories in part (a) and of the simple antisymmetric heteroclinic trajectories in part (b). The black dots mark the contributions from the structures of the NHIM equally labelled in Fig.~\ref{map}. The green curve is an interpolation of the black dots. The red stars mark points along the green curves where two branches meet. The further coordinates in part (a) are always $p_y = 0$ and $p_z = 0$. The further coordinates in part (b) are always $p_y = 0$ and $z = 0$. (Colour figure online).}
\label{pts}
\end{figure*}

\begin{figure*}
\centering
\resizebox{\hsize}{!}{\includegraphics{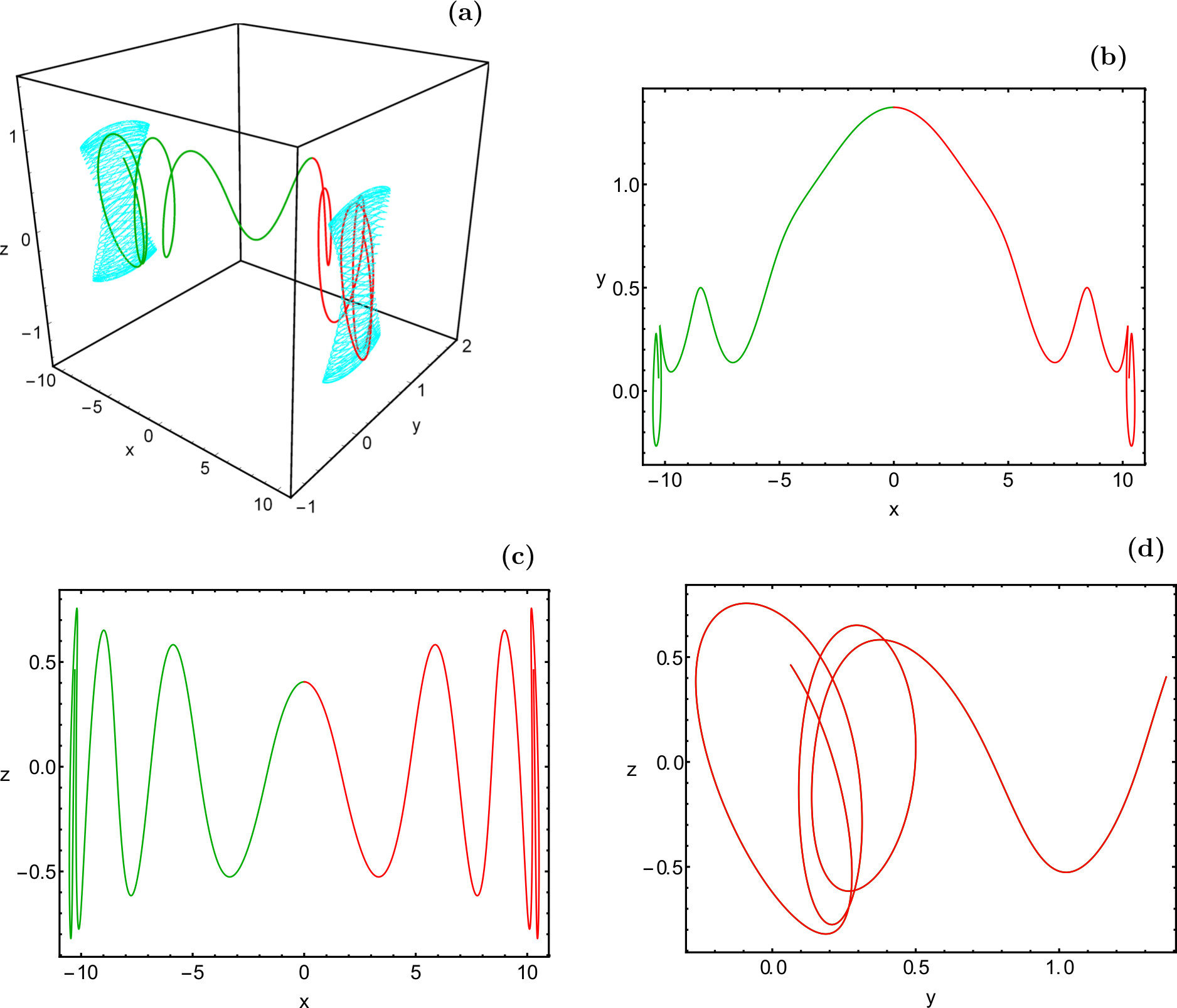}}
\caption{A simple symmetric heteroclinic trajectory connecting $i_6(L_2)$ with $i_6(L_3)$. Part (a) is a perspective view in the 3 dimensional position space $(x,y,z)$. Parts (b), (c), and (d) are the projections into the various 2 dimensional coordinate planes. The part of the trajectory with $x > 0$ (past part) is
plotted in red and the part with $x < 0$ (future part) is plotted in green. Also included, in cyan colour, are the limit sets $i_6(L_2)$ and $i_6(L_3)$ in part (a). (Colour figure online).}
\label{orb3}
\end{figure*}

\begin{figure*}
\centering
\resizebox{\hsize}{!}{\includegraphics{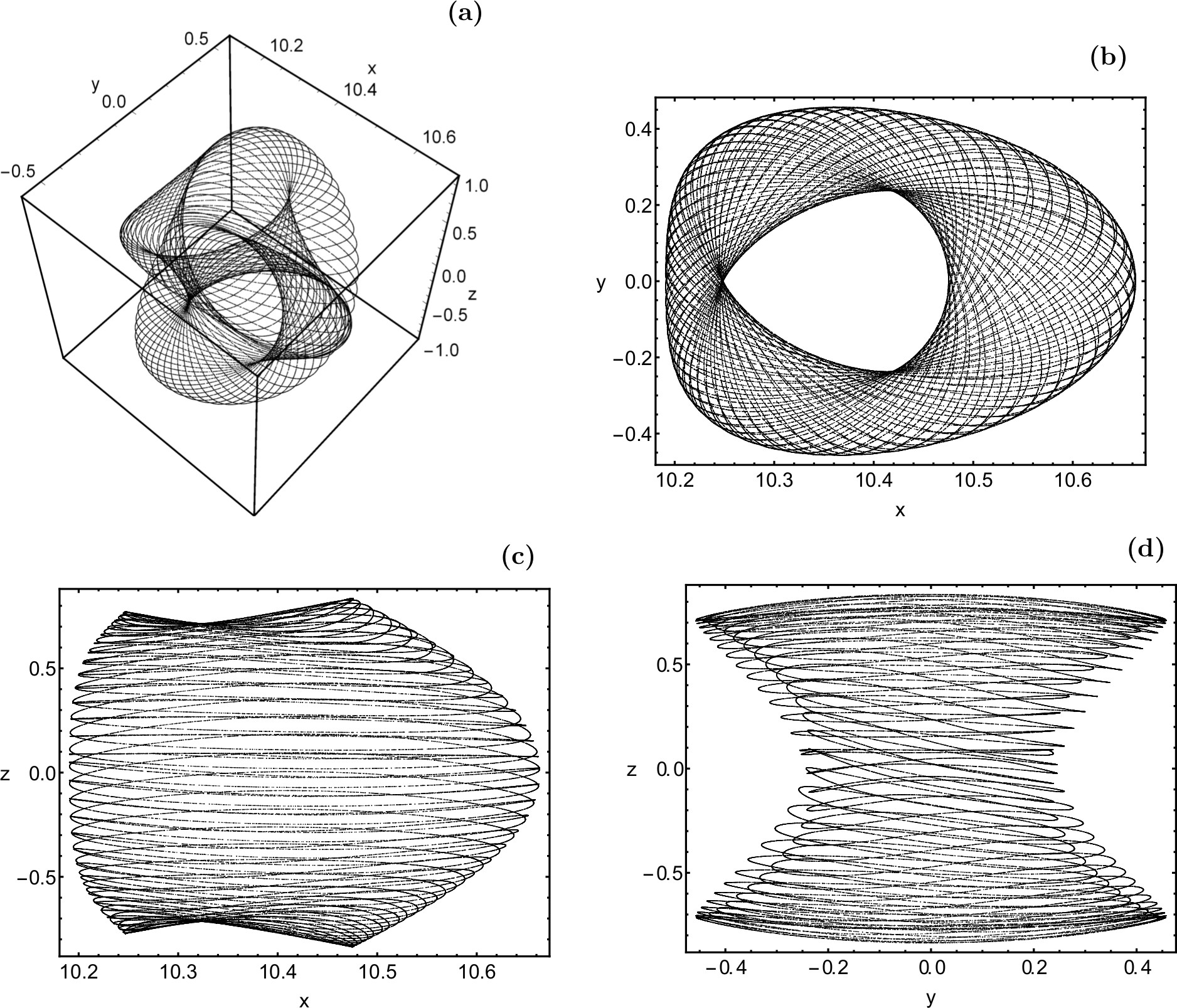}}
\caption{A magnified plot of the substructure $i_6(L_2)$ of NHIM2. Part (a) is a perspective view of a projection into the 3 dimensional position space $(x,y,z)$. Parts (b), (c), and (d) are the projections into the various 2 dimensional coordinate planes. For details on the construction of this plot see the main text.}
\label{torus}
\end{figure*}

\begin{figure*}
\centering
\resizebox{\hsize}{!}{\includegraphics{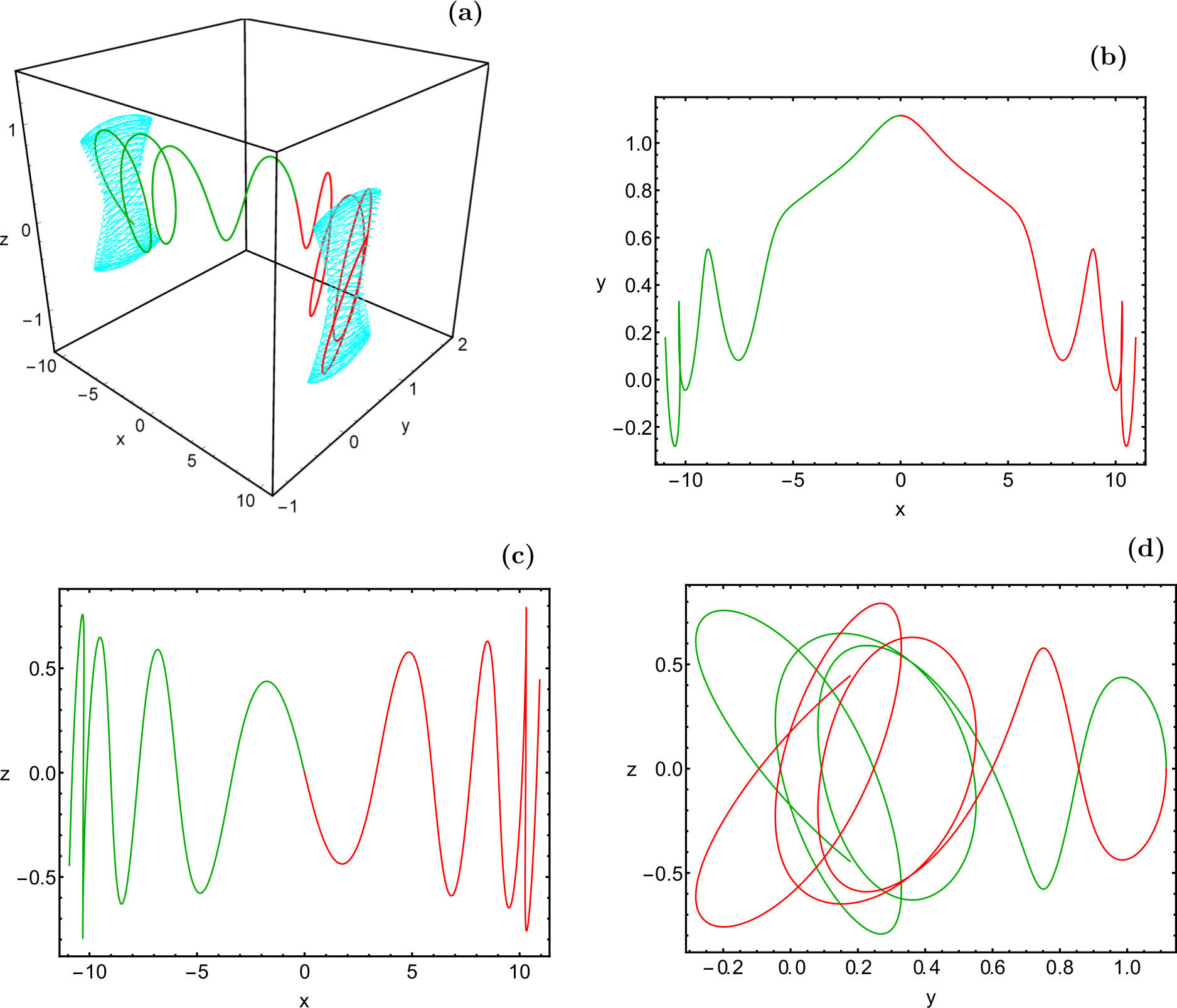}}
\caption{A simple antisymmetric heteroclinic trajectory connecting $i_6(L_2)$ with $i_6(L_3)$. Part (a) is a perspective view in the 3 dimensional position space $(x,y,z)$. Parts (b), (c), and (d) are the projections into the various 2 dimensional coordinate planes. The part of the trajectory with $x > 0$ (past part) is
plotted in red and the part with $x < 0$ (future part) is plotted in green. Also included, in cyan colour, are the limit sets $i_6(L_2)$ and $i_6(L_3)$ in part (a). (Colour figure online).}
\label{orb4}
\end{figure*}

The following considerations are the logical initial steps for our search of simple symmetric and antisymmetric heteroclinic trajectories, i.e. heteroclinic trajectories crossing the plane $x = 0$ only once and having the same symmetry properties as the periodic orbits shown in Fig.~\ref{orb1} or Fig.~\ref{orb2}, respectively. We will again use the labels class 2 and class 3 respectively for these two groups of trajectories. First, it should be clear that a single
heteroclinic trajectory going from $L_2$ to $L_3$ can never have the reflection symmetry in $y$. However, to any trajectory going from $L_2$ to $L_3$ there is the rotated trajectory going from $L_3$ to $L_2$. And we should take these two $D_z(\pi)$ related trajectories together as counterpart for a single periodic trajectory. The two rotation related heteroclinic trajectories taken together have the correct symmetry properties. Second, to obtain the desired symmetry property in $x$ it should be clear that we have to look for heteroclinic trajectories connecting equivalent (rotation symmetry related) substructures in the two NHIMs.

Let us identify simple heteroclinic trajectories from $L_2$ to $L_3$ by giving their coordinates in the moment when they cross the plane $x = 0$ in negative orientation. In the intersection point they have some value $y_0$, because of symmetry reasons (when they belong to class 2 or to class 3) they have $p_y = 0$. For the symmetric class 2 they have $p_{z,0} = 0$ and a value $z_0 \ne 0$, and for the antisymmetric class 3 they have $z_0 = 0$ and a value $p_{z,0} \ne 0$. That is, heteroclinic trajectories from class 2 are identified by giving $y_0$ and $z_0$ and heteroclinic trajectories from class 3 are identified by giving $y_0$ and $p_{z,0}$ at the moment of crossing the plane $x = 0$.

In Fig.~\ref{pts} we present the initial conditions of these symmetric heteroclinic trajectories. Part (a) gives the initial conditions of class 2 on the $(y,z)$ plane and part (b) gives the initial conditions of class 3 on the $(y,p_z)$ plane. The points are marked with labels corresponding to the substructures of the NHIMs included and labelled in Fig.~\ref{map}. Note that not all substructures lead to simple symmetric and/or antisymmetric heteroclinic connections and that some substructures of type $i_n$ and $a_n$ lead to 2 different heteroclinic connections of each symmetry class while the separatrix and some substructures in the tilted loop islands lead to four different heteroclinic connections of each symmetry class. The local branches (segments leading to the first intersections with the plane $x = 0$) of the stable and unstable manifolds of the substructures $i_4$, $i_3$, $i_2$, $i_1$, $t_1$, $t_2$ and the ones of the tilted loop orbits and of $l_v$ do not reach any point of the plane $x = 0$ with $p_y = 0$ and $z = 0$ or with $p_y = 0$ and $p_z = 0$. Therefore, we do not find corresponding simple symmetric or simple antisymmetric heteroclinic trajectories. In addition for $t_3$, $t_4$ and $t_5$ we do not find simple antisymmetric heteroclinic connections whereas the symmetric ones exist. In Fig.~\ref{map} only a small number of the substructures has been included and labelled. In total there is an infinity of further substructures and many of them lead to simple symmetric and antisymmetric heteroclinic connections. They are indicated in Fig.~\ref{pts} by the green curves. The red stars on the green curves mark the boundary points between different branches, where the sequence of labels turns its orientation and repeats labels. This means a collision of two heteroclinic trajectories, i.e. a heteroclinic bifurcation.

The symmetric connections between the two horizontal Lyapunov orbits (the trajectories presented in Fig.~\ref{pn}) can be considered limiting cases as well for class 2 as for class 3. A horizontal trajectory fulfills at the same time the defining conditions of class 2 and of class 3. These symmetric horizontal heteroclinic trajectories are the end points of the curves plotted in both parts of Fig.~\ref{pts}, i.e. in both symmetry classes. In the same figure only contributions for positive values of $z$ in part (a) or positive values of $p_z$ in part (b) are included. Because of symmetry reasons also the corresponding contributions with negative values exist. Therefore we can supplement the two parts of the figure by the vertically reflected plots and thereby in both parts the green curve turns into a closed loop with the topology of a circle. Then there are no longer any end points of the green curves.

In Figs.~\ref{orb3} and \ref{orb4} we present, as numerical examples, the symmetric and antisymmetric heteroclinic trajectories respectively in position space for the substructure $i_6$ of the NHIMs. Of course, for all plots the heteroclinic trajectories are truncated at some finite time, when they are already close to the limit sets. This holds in the past and in the future.

Because of symmetry reasons in Fig.~\ref{orb3}(d) the green segment (future segment of the trajectory) coincides exactly with the red segment (past segment of the trajectory) and is covered by the red segment and is invisible. Also included by blue colour in Fig.~\ref{orb3}(a) is a projection into the position space of the limit sets over the two saddles, which are the substructures $i_6$ of the NHIMs. In the full dimensional phase space these substructures have the topology of a 2 dimensional torus. The projection into the position space still gives an impression of this torus shape. In Fig.~\ref{orb3}a this limit set is not well resolved, therefore we repeat this limit structure over the saddle $L_2$ in better resolution in Fig.~\ref{torus}.

To produce this plot the following has been done. First 500 points on the substructure $i_6$ in Fig.~\ref{map} have been picked. All these points have $z = 0$ and have been used as initial conditions for the trajectories. Each one of these trajectories has been integrated until the next intersection with the plane $z = 0$ in the same orientation. Point sequences along these 500 trajectory segments are plotted in order to visualize the projected torus. The corresponding projected torus over the saddle $L_3$ is obtained by an application of $D_z(\pi)$ to Fig.~\ref{torus}.

The two limit sets in Fig.~\ref{orb4} coincide with the ones in Fig.~\ref{orb3} and are again given by the torus magnified in Fig.~\ref{torus}. To each heteroclinic trajectory of class 2 and of class 3 exists also the $z$ reflected heteroclinic trajectory. We do not distinguish these two trajectories.

So far we have treated the most simple heteroclinic trajectories with particular symmetry properties. There are also simple nonsymmetric heteroclinic trajectories. Just consider a heteroclinic trajectory starting on NHIM2 on the substructure $s_1$ and ending on NHIM3 on substructure $s_2$, where these two substructures are different, i.e. are not identified by an application of $D_z(\pi)$. Then it is immediately clear that this heteroclinic trajectory can not belong to the classes 1 or 2 or 3 considered so far.

\begin{figure*}
\centering
\resizebox{\hsize}{!}{\includegraphics{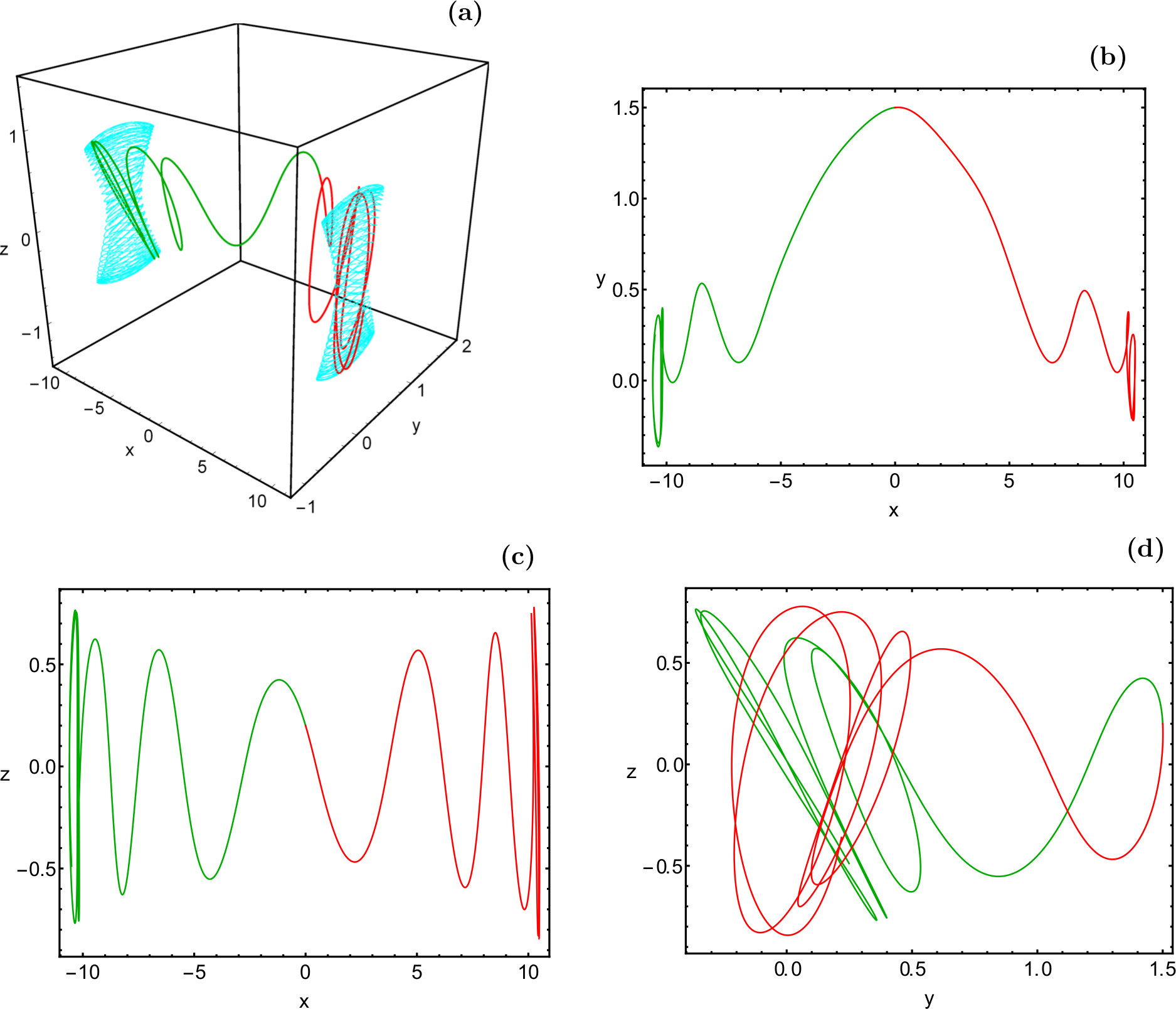}}
\caption{A simple nonsymmetric heteroclinic trajectory starting close to $i_6(L_2)$ and ending close to $i_5(L_3)$. Part (a) is a perspective view in the 3 dimensional position space $(x,y,z)$. Parts (b), (c), and (d) are the projections into the various 2 dimensional coordinate planes. The part of the trajectory with $x > 0$ (past part) is plotted in red and the part with $x < 0$ (future part) is plotted in green. Also included, in cyan colour, are the limit sets over the potential saddles. (Colour figure online).}
\label{orb5}
\end{figure*}

\begin{figure*}
\centering
\resizebox{\hsize}{!}{\includegraphics{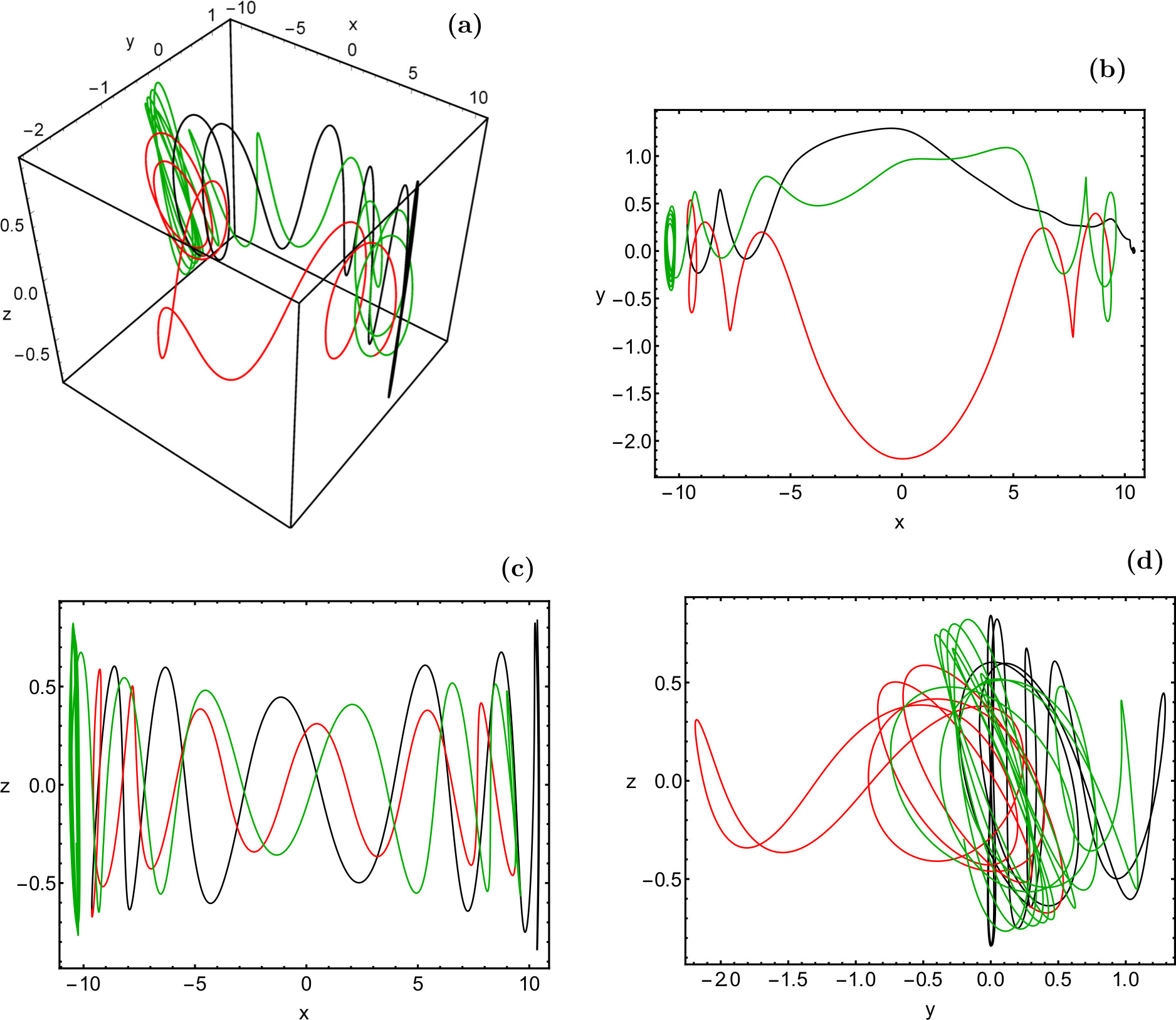}}
\caption{A nonsimple heteroclinic trajectory starting close to $l_v(L_2)$ and ending close to $i_6(L_3)$. In order to make it easier to follow the trajectory various segments are plotted in different colours. For more details see the main text. (Colour figure online).}
\label{orb6}
\end{figure*}

Fig.~\ref{orb5} is an example where the past limit set is close to $i_6(L_2)$ and the future limit set is close to $i_5(L_3)$. In part (c) we see clearly that we have a nonsymmetric trajectory. In the moment of the crossing of the plane $x = 0$ the value of $z$ is neither zero nor is it an extremal value, i.e. also $p_z$ is different from zero. We also observe that at the moment of this crossing the value of $p_y$ is rather small and the projection into the $(x,y)$ plane is close to symmetric. As a consequence, the trajectory connects substructures of the NHIMs which lie rather close in Fig.~\ref{map}. All simple heteroclinic trajectories have this property because of the following explanation.

The simple heteroclinic trajectories always have rather short trajectory segments in the central region of the bar and spend very little time in this central region. As seen in the NHIM plot of Fig.~\ref{map} the dynamics over the saddles is almost regular and this means that the partition of the available total energy between horizontal motion and vertical motion is almost constant. This energy distribution can only be changed significantly along some trajectory segment
clearly distant from the saddle region. However, also in other regions of the position space this energy transfer between horizontal and vertical motion is slow. Therefore, to obtain a significant energy transfer we need a trajectory which stays away from the saddle regions for a sufficiently long time. As we have just seen, the simple heteroclinic trajectories like the one shown in Fig.~\ref{orb5} do not do this. Therefore, these trajectories end over the saddle $L_3$ with almost the same vertical energy with which they have started over the saddle $L_2$. This explains why they connect equal or neighbouring substructures of the NHIMs. In addition, $l_v$ and the substructures close to it, like $i_1$ or $i_2$, or also the tilted loop orbits and structures close to them, like $t_1$ or $t_2$, do not contribute to the simple heteroclinic trajectories at all. In this context see again Fig.~\ref{pts}.

To get heteroclinic trajectories connecting more distant substructures of the NHIMs or having $l_v$ or the tilted loop orbits and their neighbourhoods as past or future limits these trajectories must make some extra loops away from the saddle regions. This means they must have multiple intersections with the plane $x = 0$ and can not be simple heteroclinic trajectories. As a numerical example we show in Fig.~\ref{orb6} a nonsimple heteroclinic trajectory starting close to $l_v(L_2)$ and ending over $L_3$, close to the substructure $i_6$ of NHIM3. In order not to overload the plot we did not include these limit sets into the figure. To make it easier to follow the trajectory we have cut it into 3 time segments and plotted the first segment $(t \in [0,3.2])$ in black, the second segment $(t \in [3.2, 5.1])$ in red and the third segment $(t \in [5.1,10])$ in green. Remember that in the plot we only show a finite segment of the heteroclinic trajectory which in principle runs for ever into the past and into the future.

\section{Global description of the set of simple heteroclinic trajectories}
\label{glb}

Now we consider the set of all simple heteroclinic trajectories connecting NHIM2 and NHIM3, let us call this set $\tilde{S}$. For the moment, we concentrate on the ones which have NHIM2 as past limit and NHIM3 as future limit. Each one of these trajectories intersects only once the intersection surface $R$ defined by the condition $x = 0$. Therefore, we can represent each element of $\tilde{S}$ by a point in $R$. Let us call this corresponding set of intersection points $S$. There is a 1:1 relation between trajectories from $\tilde{S}$ and points from $S$. First let us discuss the dimension of $S$. We still consider a single value $E$ of the total energy only. The dimension of the corresponding energy shell in the phase space is 5. The dimension of the intersection surface $R$ is 4, as coordinates in $R$ we naturally use $y$, $z$, $p_y$, and $p_z$. The dimension of NHIM1 and NHIM2 for fixed energy is 3. The dimension of the stable and unstable manifolds of the NHIMs is 4. The dimension of the intersections between the stable and unstable manifolds of the NHIMs and $R$ is 3. Let us call these intersections $M_2$ and $M_3$ for the intersection of the local branch of the unstable manifold of NHIM2 and the local branch of the stable manifold of NHIM3, respectively. Local means here that we only consider first intersections between $R$ and trajectories running along the stable and unstable manifolds and we ignore possible later additional intersections. Simple heteroclinic trajectories are then given by intersections between $M_2$ and $M_3$. In a nondegenerate case this is the transverse intersection between two 3 dimensional sets located in a 4 dimensional embedding set. This intersection is the set $S$ defined before. We can interpret it as the primary heteroclinic intersection set between NHIM2 and NHIM3. And according to the previous considerations its dimension is 2.

Next we need the argument that the topology of $S$ is the one of a 2 dimensional sphere. We start our considerations in the 5 dimensional energy shell in the phase space. There the NHIMs are 3 dimensional surfaces and have the topology of a 3 dimensional sphere $S^3$. For the stable and the unstable manifolds of NHIMs we have the foliation theorem which shows that the internal structure of these manifolds is essentially a Cartesian product of the NHIM and a line. Then the transverse intersection between the stable or the unstable manifold and the hypersurface $R$ reproduces a continuous image of the NHIM and therefore it also has the topology
of $S^3$. This holds for the unstable manifold of NHIM2 and also for the stable manifold of NHIM3. Then transverse nonempty intersections of these two manifolds within $R$ are nonempty transverse intersections between 2 copies of $S^3$ embedded in the 4 dimensional manifold $R$. And this intersection (i.e. $S$) is 2 dimensional and has the topology of a 2 dimensional sphere $S^2$.

The intersection of $S$ with the plane ($p_y = 0$, $p_z =0$) is the green curve in Fig.~\ref{pts}a representing simple symmetric heteroclinic trajectories together with its $z$ reflected mirror image. Now it should no longer be surprising that the intersection between a surface $S$ with the topology of $S^2$ and a plane gives a curve of the topology of a circle. Let us call this curve $C_s$. We can imagine the curve $C_s$ as the curve on $S^2$ with constant azimuth angle 0 and $\pi$. We can define this azimuth angle as $\phi = \arctan ( p_z / z )$. And the intersection of $S$ with the plane ($p_y = 0$, $z = 0$) is the green curve in Fig.~\ref{pts}b representing simple antisymmetric heteroclinic trajectories together with its $p_z$ reflected mirror image. It is again a curve with the topology of a circle. Let us call this curve $C_a$. We can imagine the curve $C_a$ as the curve on $S^2$ with constant azimuth angle $\pm \pi/2$. The two horizontal simple heteroclinic trajectories shown in Fig.~\ref{pn} are the intersection between $S$ and the $y$ axis, i.e. they are the two points on $S$ fulfilling simultaneously $z = 0$, $p_y = 0$ and $p_z = 0$. They are the two intersection points between the curves $C_s$ and $C_a$. We can imagine these two points as the two poles of $S^2$. The nonsymmetric simple heteroclinic trajectories fill the whole rest of $S$ not belonging to the two curves $C_s$ and $C_a$. Let us call this complement $C_n$. We can imagine $C_n$ as all points on $S^2$ with an azimuth angle which is not an integer multiple of $\pi/2$.

Next we can imagine that we cover $S$ by two different systems of 1 dimensional curves. We call these systems of curves $SC_2$ and $SC_3$. Elements of these two sets are labelled by the substructures of the NHIMs. A curve $SC_2(s_n)$ from the set $SC_2$ contains points on $S$ which have the substructure $s_n$ of NHIM2 as past limit set. And a curve $SC_3(s_k)$ from the set $SC_3$ contains points on $S$ which have the substructure $s_k$ of NHIM3 as future limit set. Note that not
all existing substructures on the NHIMs have corresponding curves in the sets $SC_2$ and $SC_3$, this happens for example for the substructures $i_1$, $i_2$, $i_3$, $i_4$, $t_1$, $t_2$, $l_v$ and for the tilted loop orbits. It principle it can be allowed that some of these curves consist of various connected components. The intersection points between $SC_2(s_n)$ and $SC_3(s_k)$ represent the simple heteroclinic trajectories going from the substructure $s_n$ of NHIM2 to the substructure $s_k$ on NHIM3, while crossing the plane $x = 0$ only once. The intersections between $SC_2(s_n)$ and $SC_3(s_k)$, where $s_n$ and $s_k$ are the same substructures on NHIM2 and NHIM3 (i.e. when $s_n$ on NHIM2 is transformed into $s_k$ on NHIM3 by $D_z(\pi)$) are exactly the points along the curves $C_s$ and $C_a$, i.e. they represent the simple symmetric and antisymmetric heteroclinic trajectories. Curves from the set $SC_2$ really have intersections only with a
part of the curves from $SC_3$. Because of the discrete symmetries of the system the number of intersection points between a curve from $SC_2$ with a curve from $SC_3$ can be either 0 or 4 or 8 or 12 or 16, when we count all symmetry related copies of simple heteroclinic trajectories.

For the simple heteroclinic trajectories going from NHIM3 to NHIM2 we have an equivalent sphere. These two equivalent spheres are transformed into each other by $D_z(\pi)$.

We have given numerical examples for the energy $E = -3200$ only. The qualitative description of $S$ is similar for all energy values a little higher than the saddle energy. Of course, it depends on the energy value exactly which substructures from NHIM2 and NHIM3 are connected by symmetric or by antisymmetric or by nonsymmetric simple heteroclinic trajectories.

Besides the surface $S$ representing the primary heteroclinic intersection surface and consisting of simple heteroclinic trajectories there is an infinity of other heteroclinic intersection surfaces representing more complicated heteroclinic trajectories making additional loops and intersecting the surface $x = 0$ several times. Such heteroclinic trajectories can connect substructures from NHIM2 and NHIM3 which are not connected by simple heteroclinic trajectories. Remember the example shown in Fig.~\ref{orb6}.

The description given here relies heavily on the discrete symmetries of the system. But it should be clear that the qualitative picture remains valid under small perturbations of the symmetry.

\section{Interpretation of described periodic orbits and heteroclinic trajectories as x1 orbits}
\label{x1}

By looking over the various figures showing heteroclinic trajectories and periodic orbits close to the heteroclinic tangle we note the following common feature: All these trajectories are confined to a narrow strip in $y$ (approximately between -1.2 and +1.2) whenever $|x| > 5$. In addition, when $|x| < 5 $ then only a small relative fraction of these trajectories enters the interior of the nucleus. These trajectories have a large density in a narrow shell around the nucleus. In this sense this set of trajectories traces out the outer parts of the bar together with a shell around the nucleus. The individual trajectories from the neighbourhood of this set are rather unstable. However, when we perturb some trajectory from this set, then it switches to a similar trajectory of the same set. In this sense this whole set of trajectories is rather robust, dynamically and also structurally. All heteroclinic trajectories run along the bar axis (which
coincides with the $x$-axis), with small values of $y$, while they stay rather close to the horizontal plane $(x,y)$, i.e. they only make small oscillations in $z$ direction $(|z| < 1)$. This means that all heteroclinic trajectories are excellent candidates of x1 type of orbits, which support the barred structure of the galaxy.

\begin{figure}
\begin{center}
\includegraphics[width=\hsize]{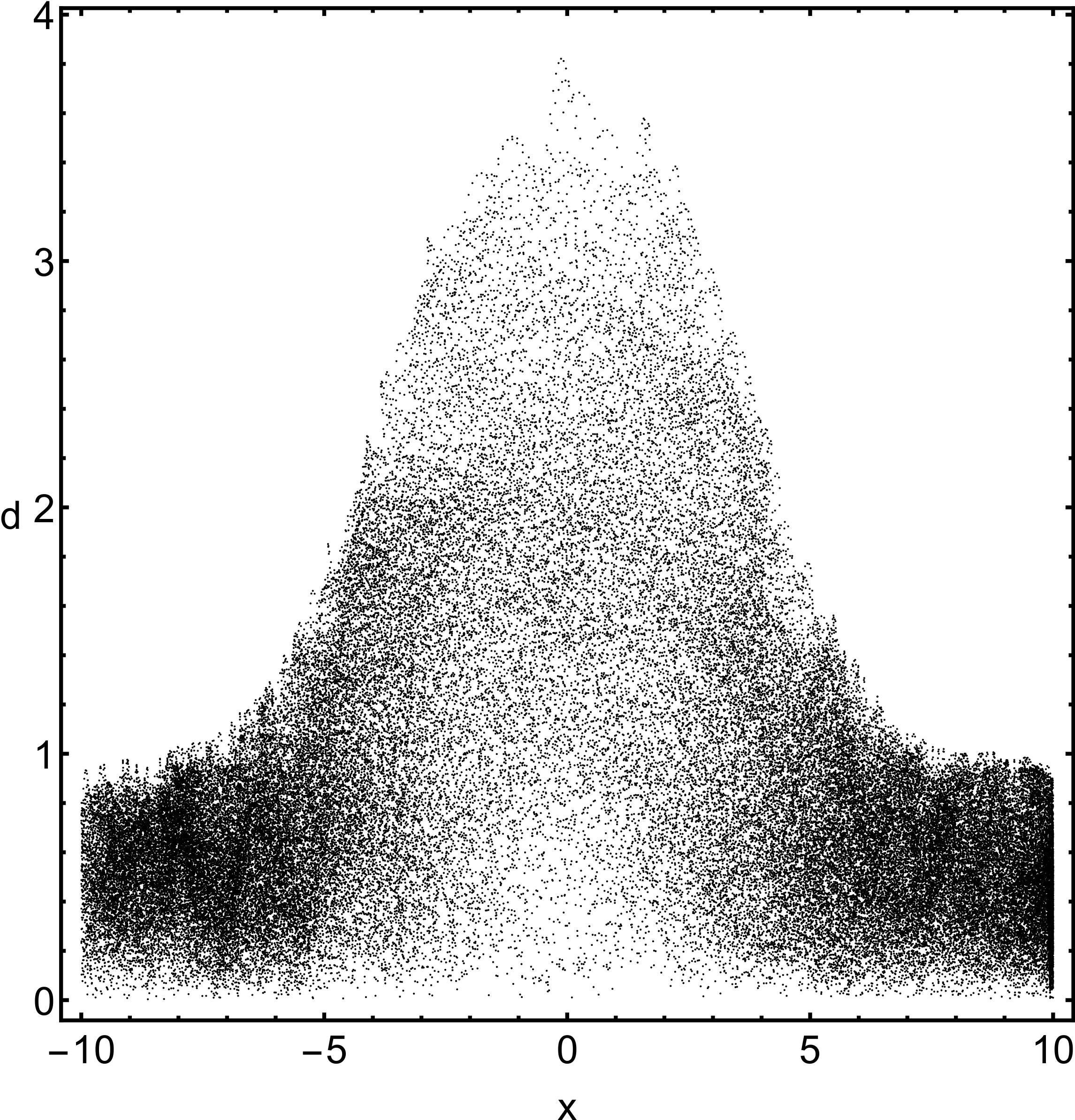}
\end{center}
\caption{Probability distribution of trajectories running close to the unstable manifold of NHIM2, plotted in the cylinder coordinates $x$ and $d = \sqrt{y^2 + z^2}$. For more details see the main text.}
\label{dist}
\end{figure}

Fig.~\ref{dist} has been constructed to demonstrate the mentioned distribution in an additional form. We have introduced cylindrical coordinates with the $x$ axis as the cylinder axis and the cylindrical radius $d = \sqrt{y^2 + z^2}$. Next we have initiated 1000 trajectories near the NHIM2 and let them run to the interior region, close to the inner branch of the unstable manifold of NHIM2. We let these trajectories run until a time $t = 10$. The figure shows the density of this collection of trajectories over the $(x,d)$ plane. We obtain a very high density in the two outer parts of the bar where however the density is relatively moderate in the direct neighbourhood of the cylinder axis. We have a small density in the interior of the nucleus and a larger density in a shell around the nucleus. This plot gives a good impression how the unstable manifolds of the saddle NHIMs and their heteroclinic tangle confine the bar and the nucleus. In a real galaxy, stars can escape from this confinement by close encounters causing a change of momentum and energy.

\begin{figure*}
\centering
\resizebox{0.8\hsize}{!}{\includegraphics{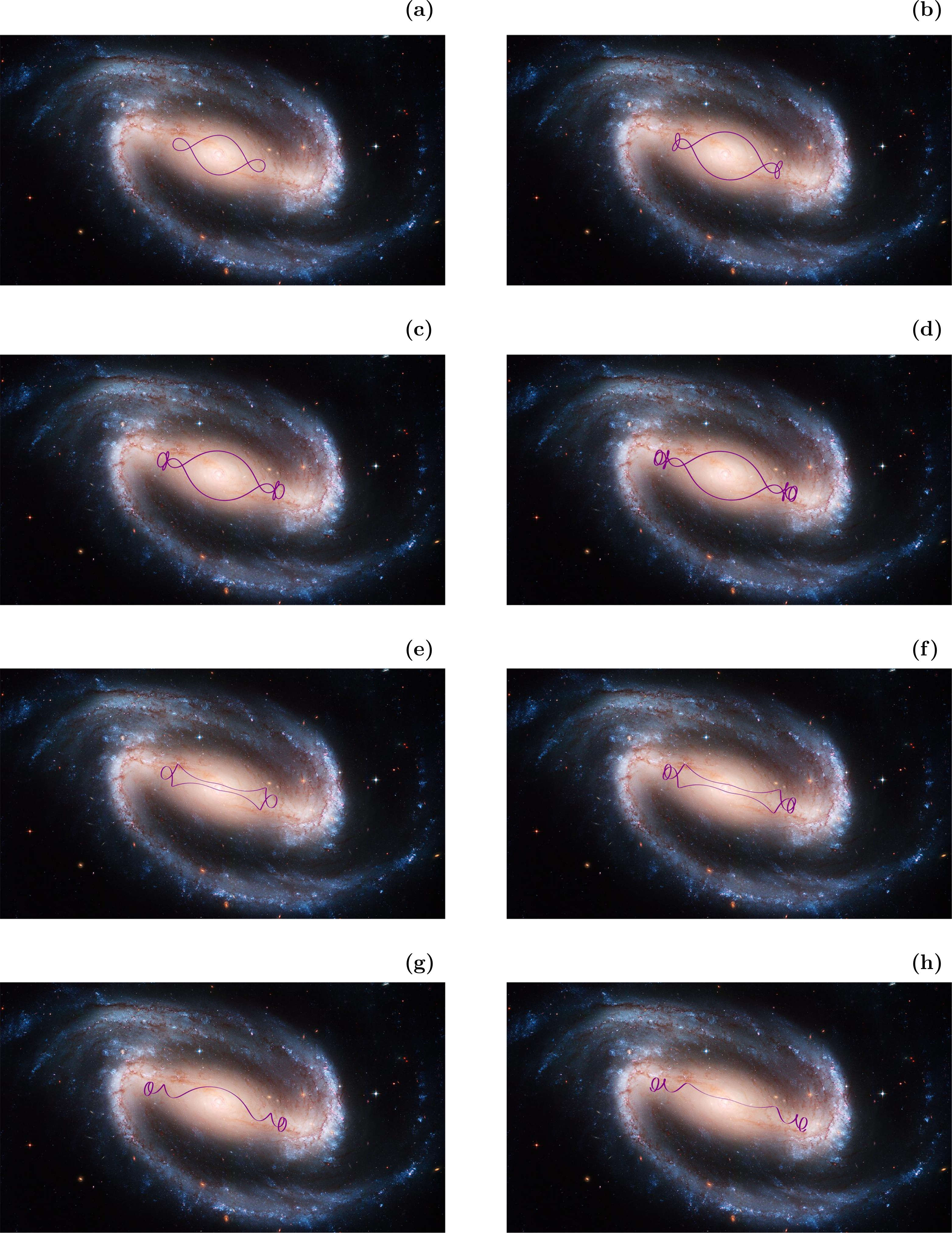}}
\caption{In the eight panels, several periodic orbits and two simple horizontal heteroclinic trajectories (in purple colour) have been rotated into the appropriate scale and direction of view and have been inserted into a real image of the barred spiral galaxy NGC 1300. The figure shows how these types of trajectories populate and shape the bar and the neighbourhood of the nucleus. The trajectories included in the parts (a), (b), (c), (d), (e), (f) are the ones shown earlier in Figs.~\ref{orbs}a, \ref{orbs}b, \ref{orbs}c, \ref{orbs}d, \ref{orbs}e, \ref{orbs}f, respectively, while the trajectories included into parts (g) and (h) are the two heteroclinic trajectories from Fig.~\ref{pn}. (Colour figure online).}
\label{sml}
\end{figure*}

\begin{figure*}
\centering
\resizebox{0.8\hsize}{!}{\includegraphics{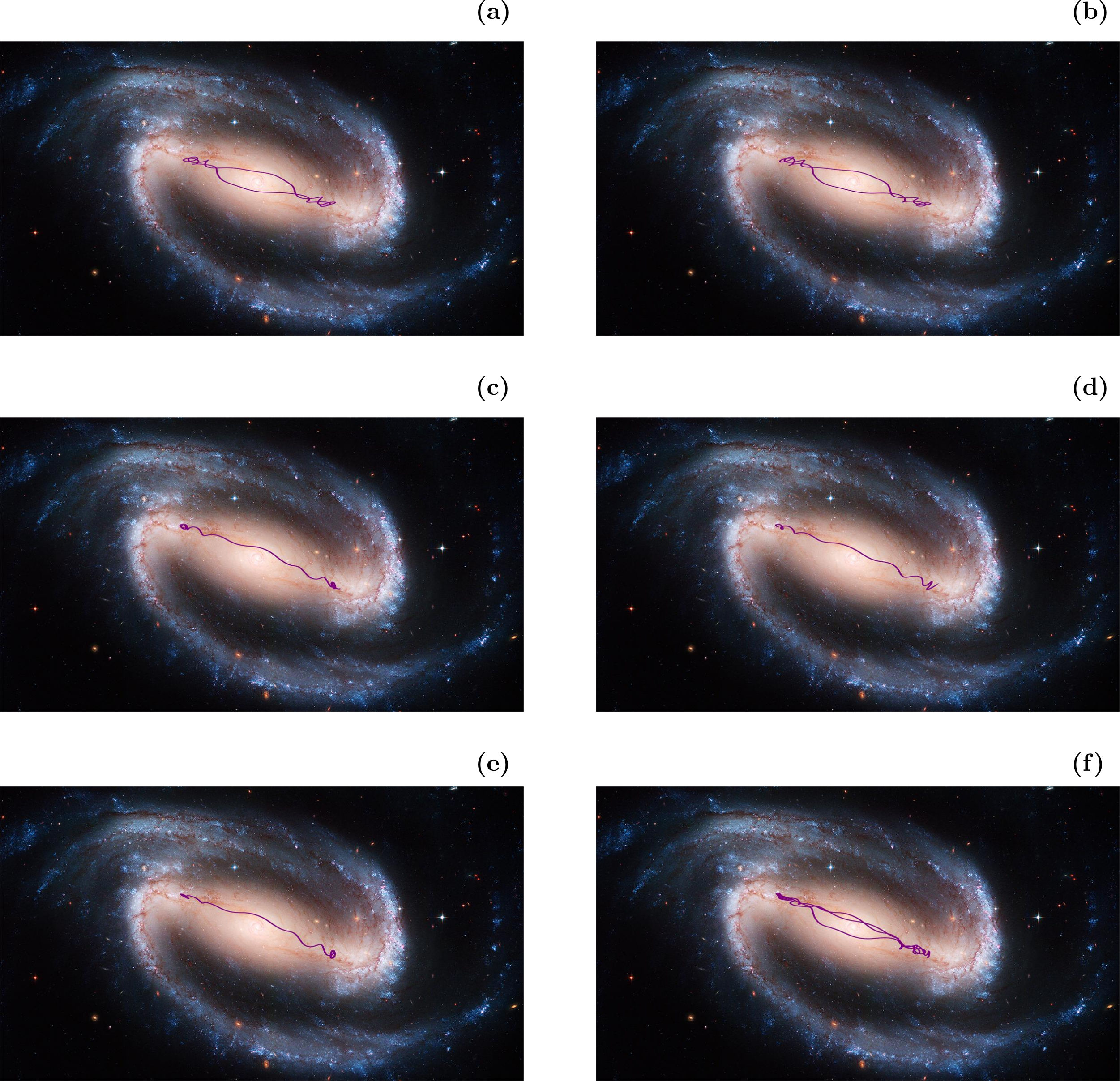}}
\caption{In the six panels, two periodic orbits and 4 heteroclinic orbits have been scaled and rotated into the appropriate perspective and have been inserted into a real image of NGC 1300. These trajectories included in parts (a), (b), (c), (d), (e) and (f) are the ones shown earlier in Figs.~\ref{orb1}, \ref{orb2}, \ref{orb3}, \ref{orb4}, \ref{orb5}, and \ref{orb6}, respectively. Again, the figure makes evident how this type of trajectories populates and shapes the bar and the neighbourhood of the nucleus. (Colour figure online).}
\label{sml2}
\end{figure*}

Next let us have a look at relevant time scales. The short, simple periodic orbits (like the ones shown in Fig.~\ref{orbs}) have periods of a few of our dimensionless time units and the time unit corresponds to approximately 100 million years, i.e. lies in the order of magnitude of the rotation period of the bar. In this sense, when a trajectory remains near the heteroclinic tangle for several time units then it remains in the bar region for several rotation periods. That is, such trajectories belong to the set of trajectories which populate the bar for some time, form and stabilize the bar. They behave like x1 orbits. For the energy $E = -3200$ we did not find any dynamically stable periodic x1 orbits, but as just explained the heteroclinic tangle seems to take over the job which usually is attributed to the periodic x1 orbits, namely to shape and stabilize the bar structure.

The heteroclinic tangle certainly is not uniformly hyperbolic. This should already be clear from the almost integrable internal dynamics of the NHIMs. That is, we have a mixed phase space. Usually in homoclinic/heteroclinic tangles with mixed phase spaces, we find stable periodic orbits, at least of high periods. Therefore, it would be no surprise for us, if also in our present heteroclinic tangle stable periodic orbits of high period would exist.

Now let us check how the calculated periodic and heteroclinic trajectories fit into the real galaxy NGC 1300. Remember, that the parameters of our potential model are chosen to fit the properties of this particular galaxy. To construct Fig.~\ref{sml} we have rotated 6 horizontal periodic orbits (the ones from Fig.~\ref{orbs}) and 2 simple horizontal heteroclinic trajectories (the ones from Fig.~\ref{pn}) into the appropriate direction of view and have included them, using the correct scale, into a real image of the barred galaxy NGC 1300. According to plate 10 in \citet{BT08} the semi-major axis of NGC 1300 is about 10 kpc. Using this as a scale measure we created a tilted frame of reference, so as to appropriately fit all trajectories on top of the real image, at the correct size and position.

Fig.~\ref{sml} contains horizontal trajectories only because for them it is easy to imagine how they are running in the full 3 dimensional position space. In contrast in the 6 panels of Fig.~\ref{sml2} we present an analogous plot for the 6 trajectories already shown in the Figs.~\ref{orb1}, \ref{orb2}, \ref{orb3}, \ref{orb4}, \ref{orb5}, \ref{orb6}, respectively. These trajectories explore all 3 degrees of freedom and to understand well their motion in these pictures of the real galaxy we need to consult the projections of these trajectories into the various coordinate planes as given in the mentioned previous plots. From Figs. \ref{sml} and \ref{sml2} it becomes evident how the periodic orbits and the heteroclinic trajectories populate the bar and the neighbourhood of the nucleus and how they form the skeleton of the bar and the surrounding of the nucleus. At the same time these combined plots indicate that our dynamical model is realistic for barred galaxies with features similar to NGC 1300.

Many barred galaxies have dust stripes in their bars and these lines start near the saddle point (connection between bar and outer spirals) and they run mainly in
longitudinal direction of the bar, but not along the symmetry axis of the bar, they are shifted to one side. When we compare these dust lines with Figs. 13a and 13b in \citetalias{JZ16a} or Fig.~9a in \citetalias{JZ16b}, then it becomes evident that these dust stripes run along the local segments of the inner branches of the unstable manifolds of the saddle NHIMs. Good examples of real galaxies showing these patterns are the following NGC numbers: 613, 1097, 1300, 1365, 1530, 4303, 5236, 5921, 6221, 6907, 6951, 7552 \citep[see e.g.,][]{KB19}. The relation of these patterns to the NHIM properties might be understood along the following arguments: Many barred galaxies have an inflow of gas and dust from the outer parts to the interior. Then it is obvious that this gas and dust first approaches the saddle from the outside along the local segments of the outer branches of the stable manifolds of the saddle NHIMs. Next, it flows over the saddle and then continues along the inward going branches of the unstable manifolds. In this sense, the dust stripes visualise the projection into the position space of the local segments of the inner branches of the unstable saddle manifolds. Of course, a part of this gas and dust which has come close to the saddle points, returns to the outside and leaves along the outer branches of the unstable saddle manifolds. Also this outer dust pattern is clearly visible in some real barred galaxies, good examples are the NGC numbers 1097, 1300, 5236, 6951, 7479.

\section{Discussion and conclusions}
\label{disc}

The NHIMs of codimension 2 sitting over the two index-1 saddle points of the effective galactic potential together with their stable and unstable manifolds direct the global orbital dynamics of the whole system to a large extent. Here also the nearby periodic orbits having a similar shape contribute. In previous publications we have already explained how the outer branches of the unstable manifolds determine the structure of outer rings or spirals (it should be mentioned that in the case of rings formed by the outer branches of the unstable manifolds there are in addition spirals in the outer part of the disk caused by other mechanisms not related to the NHIMs and their invariant manifolds). In the present article we explain how the inner branches and the corresponding heteroclinic connections are related to the bar and to a neighbourhood of the nucleus. In this sense, the important visible structures in the interior parts of barred galaxies are directly related to the projection into the $(x,y,z)$ position space of the inner branches of unstable manifolds of saddle NHIMs. Thereby the barred galaxy is the most beautiful example, known so far, to show the structure of these mathematical objects directly in position space and easily accessible to observations.

Figs.~\ref{sml} and \ref{sml2} demonstrate how the important periodic orbits, close to the heteroclinic tangle and the heteroclinic trajectories themselves, fit into the observed structure of NGC 1300. We consider this an important confirmation of our model potential, where the part describing the bar has been introduced by us as a simpler alternative to the long established standard Ferrers' triaxial model \citep{F77}. Figs.~\ref{sml} and \ref{sml2} of the present article together with Fig.~13 from \citetalias{JZ16b} give a rather complete picture how the NHIMs and the corresponding stable and unstable manifolds are the dominating subsets for the formation of structures in the system.

From the dynamical system theory point of view, we made an important progress in the detailed understanding of the primary heteroclinic intersection surface, i.e. the one representing simple heteroclinic trajectories. This 2 dimensional surface has the form of a sphere $S^2$ where the poles represent the two horizontal heteroclinic trajectories and the circle of azimuth angle 0 and $\pi$ on one hand and the circle of azimuth angle $+\pi/2$ and $-\pi/2$ on the other hand represent the symmetric and antisymmetric heteroclinic trajectories respectively. The mentioned symmetry properties refer to the $z$ motion relative to the $x$ motion. The heteroclinic trajectories with symmetry always connect equivalent substructures on the two NHIMs. The rest of the simple heteroclinic trajectories connect different substructures on the 2 NHIMs, i.e. substructures which are not identified by the discrete symmetries of the system. This description should also hold for other systems with the same discrete symmetries and many of its qualitative features should survive small perturbations of the symmetry. To our knowledge such a detailed description of the primary heteroclinic intersection surface between two NHIMs of 3-dof systems is new. Moreover, it was also shown that heteroclinic trajectories are excellent candidates of x1 types orbits which support the barred structure and geometry of the galaxy.

For numerically integrating the equations of motion we used a Bulirsch-Stoer routine in standard version of \verb!FORTRAN 77! \citep[e.g.,][]{PTVF92}, with double precision. The relative error regarding the conservation of the orbital energy was of the order of $10^{-14}$, using a fixed time step equal to 0.001 and a Quad-Core i7 vPro 4.0 GHz processor. All the graphics of the paper have been constructed by using the 11.3 version of the Mathematica$^{\circledR}$ software \citep{W03}.

\section*{Acknowledgments}

One of the authors (CJ) thanks DGAPA for financial support under grant number IG-100819. The authors would like to thank the anonymous referee for all the apt suggestions and comments which improved both the quality and the clarity of the paper.

\bsp
\label{lastpage}


\begin{thebibliography}{}

\bibitem[\protect\citeauthoryear{Athanassoula et al.}{2009a}]{ARGM09} Athanassoula E., Romero-G\'{o}mez M., Masdemont J.J., 2009a, MNRAS, 394, 67

\bibitem[\protect\citeauthoryear{Athanassoula et al.}{2009b}]{ARGBM09} Athanassoula E., Romero-G\'{o}mez M., Bosma, A., Masdemont J.J., 2009b, MNRAS, 400, 1706

\bibitem[\protect\citeauthoryear{Athanassoula et al.}{2010}]{ARGBM10} Athanassoula E., Romero-G\'{o}mez M., Bosma A., Masdemont J.J., 2010, MNRAS, 407, 1433

\bibitem[\protect\citeauthoryear{Binney \& Tremaine}{2008}]{BT08} Binney J., Tremaine S., 2008, Galactic Dynamics, 2nd edn. Princeton Univ. Press Princeton

\bibitem[\protect\citeauthoryear{Ferrers}{1877}]{F77} Ferrers N.M., 1877, Q. J. Pure Appl. Math., 14, 1

\bibitem[\protect\citeauthoryear{Gonzalez et al.}{2014}]{GDJ14} Gonzalez F., Drotos G., Jung C., 2014 J. Phys. A: Math. Theor., 47, 045101

\bibitem[\protect\citeauthoryear{Jung \& Zotos}{2015}]{JZ15} Jung Ch., Zotos, E.E., 2015, PASA, 32, e042

\bibitem[\protect\citeauthoryear{Jung \& Zotos}{2016a}]{JZ16a} Jung Ch., Zotos, E.E., 2016a, MNRAS, 457, 2583 (Part I)

\bibitem[\protect\citeauthoryear{Jung \& Zotos}{2016b}]{JZ16b} Jung Ch., Zotos, E.E., 2016b, MNRAS, 463, 3965 (Part II)

\bibitem[\protect\citeauthoryear{K\"onig \& Binnewies}{2019}]{KB19} K\"onig M., Binnewies S., 2019, Bildatlas der Galaxien, 2. Auflage Editorial: Kosmos Verlag, Stuttgart

\bibitem[\protect\citeauthoryear{Lyapunov}{1907}]{L07} Lyapunov A.M., 1907, Ann. Fac. Sci. Toulouse 9, 203

\bibitem[\protect\citeauthoryear{Lyapunov}{1949}]{L49} Lyapunov A.M, 1949, Annals of Mathematical Studies, Vol. 17

\bibitem[\protect\citeauthoryear{Pfenniger}{1984}]{P84} Pfenniger D., 1984, A\&A 134, 373

\bibitem[\protect\citeauthoryear{Press}{1992}]{PTVF92} Press H.P., Teukolsky S.A, Vetterling W.T., Flannery B.P., 1992, Numerical Recipes in FORTRAN 77, 2nd Ed., Cambridge Univ. Press, Cambridge, USA

\bibitem[\protect\citeauthoryear{Romero-G\'{o}mez et al.}{2006}]{RGMA06} Romero-G\'{o}mez M., Masdemont J.J., Athanassoula E., Garc\'{i}a-G\'{o}mez C., 2006, A\&A, 453, 39

\bibitem[\protect\citeauthoryear{Romero-G\'{o}mez et al.}{2007}]{RGAM07} Romero-G\'{o}mez M., Athanassoula E., Masdemont J.J., Garc\'{i}a-G\'{o}mez C., 2007, A\&A, 472, 63

\bibitem[\protect\citeauthoryear{Tsoutsis et al.}{2008}]{TEV08} Tsoutsis P., Efthymiopoulos C., Voglis N., 2008, MNRAS, 387, 1264

\bibitem[\protect\citeauthoryear{Tsoutsis et al.}{2009}]{TKEC09} Tsoutsis P., Kalapotharakos C., Efthymiopoulos C., Contopoulos G., 2009, A\&A, 495, 743

\bibitem[\protect\citeauthoryear{Voglis et al.}{2006}]{VTE06} Voglis N., Tsoutsis P., Efthymiopoulos C., 2006, MNRAS, 373, 280

\bibitem[\protect\citeauthoryear{Wiggins}{1994}]{W94} Wiggins S., 1994, Normally Hyperbolic Invariant Manifolds in Dynamical Systems, Berlin: Springer Verlag

\bibitem[\protect\citeauthoryear{Wolfram}{2003}]{W03} Wolfram S., 2003, The Mathematica Book. Wolfram Media, Champaign

\bibitem[\protect\citeauthoryear{Zotos \& Jung}{2018}]{ZJ18} Zotos E.E., Jung, Ch., 2018, MNRAS, 473, 806 (Part III)

\end{thebibliography}
\end{document}